# Aberration correction in epi-fluorescence microscopy using unknown speckle illumination

EVOLENE PREMILLIEU,[1,*] ANTONIO M. CARAVACA-AGUIRRE,[1] SIMON LABOUESSE,[1] KRISTINA IRSCH,[2] AND RAFAEL PIESTUN[1]

[1]*Department of Electrical Computer and Energy Engineering, University of Colorado Boulder, UCB 425, Boulder, CO 80309, USA*
[2]*Vision Institute - CNRS, INSERM, Sorbonne University, 17 Rue Moreau, 75012 Paris, France*
*\*evolene.premillieu@colorado.edu*

**Abstract:** Diffraction-limited imaging in epi-fluorescence microscopy remains a challenge when sample aberrations are present or when the region of interest rests deep within an inhomogeneous medium. Adaptive optics is an attractive solution albeit with limited field of view and requiring relatively complicated systems. Alternatively, reconstruction algorithms have been developed over the years to correct for aberrations. Unfortunately, purely post-processing techniques tend to be ill-posed and provide only incremental improvements in image quality. Here, we report a computational optical approach using unknown speckle illumination and matched reconstruction algorithm to correct for aberrations and reach or surpass diffraction limited resolution. The data acquisition is performed by shifting an unknown speckle pattern with respect to the fluorescent object. The method recovers simultaneously a high-resolution image, the point spread function of the system that contains the aberrations, the speckle illumination pattern, and the shift positions.

## 1. Introduction

Overcoming optical aberrations is the raison d'être of optical design. Unfortunately, changes in the medium connecting the object (sample) and the photosensitive device (e.g. camera) render the best optical designs moot. Aberrations can occur as a result of focusing the optics inside a medium and through interfaces. Alternatively, the medium might change as a result of dynamic modifications occurring in live tissue, for instance, due to blood circulation. Environmental changes such as temperature or pressure might affect the medium as well. Adaptive Optics (AO) elegantly addresses this issue utilizing deformable mirrors to compensate the wavefront distortions due to the optical imperfections from an intermediate medium. While extremely successful, AO has its limitations, being sensitive to deep samples, to positioning errors, to color dispersion, and presenting a limited field of view restricting imaging to isoplanatic patches [1]. Hence, AO results in relatively complex and bulky systems [2]. Structured Illumination (SI), on the other hand, collects multiple images with known illuminations to further sample the high spatial frequencies in the object to recover high resolution images [3–5]. However, traditional sinusoidal SI is strongly limited in the presence of aberrations because the reconstruction algorithms require precise knowledge of these patterns [6]. Alternatively, SI implemented with random speckle illumination is often not affected by aberrations because, in their presence, speckle patterns transform into a different speckle pattern with similar statistical characteristics [7].

Since the invention of the laser, speckle patterns have been used for precision sensing and imaging [8–10]. However, with the advent of high information content electronic imaging and ever-increasing computation capabilities, novel computational/optical techniques have been proposed. SI microscopy using a series of different unknown speckle patterns has been demonstrated under the assumption that the average illumination is uniform over the object [11–13]. Speckle pattern illumination has been used lately to push the limits of resolution in a variety of experimental conditions [14–18]. They rely on optimization-based algorithms to recover high-resolution images of the object using prior knowledge of the system, such as the average speckle size of the illumination pattern [18], the scan positions [15], or the system's shift invariance [14]. Unfortunately, these techniques are not directly applicable to epi-fluorescence microscopy. Instead, they use a high numerical aperture (NA) lens in transmission or an adjacent diffuser to generate tiny (smaller than the diffraction limit of the collection optics) speckle patterns.

Here, we present a method to reconstruct a high-resolution diffraction-limited image and the overall aberrated Point Spread Function (PSF) of the system in epi-fluorescence microscopy. Instead of using multiple unknown speckle patterns, we scan a single speckle illumination with respect to the sample. The algorithm does not require any prior knowledge of the illumination pattern, the PSF of the system, or the scan positions. In particular, for the reconstruction, we use a gradient descent algorithm with block convex criterion [12]. We name the method Blind Speckle Illumination Fluorescence Imaging (B-SIFI). One advantage of B-SIFI is that it can be implemented on practically any imaging system as long as the speckle illumination pattern can be scanned with respect to the sample.

## 2. Theory

In a linear shift-invariant system, the acquired images can be expressed mathematically as the convolution between the PSF and the object times the illumination. Hence, the images obtained using structured illumination can be expressed as:

$$y(r,x) = \int h(r-r')\rho(r')I(r'-x)dr',$$

where $y$ represents the image, $h$ the PSF of the imaging system, $\rho$ is the object and $I$ is the structured illumination. $r$ is the position and $x$ is the relative shift increment between object and illumination. Typically, the PSF and the illumination are known a priori, but in many cases such as in the presence of sample-induced aberrations, they could lead to image reconstruction degradation and artifacts. Here, we assume that both the PSF and the illuminations are unknown, and we only measure $y(r,x)$. To recover all unknown variables, we rewrite the previous equation as a minimization problem of the functional $J$:

$$J = \sum_i \left\| y^i - h^k \otimes (\rho^k . I_i^k) \right\|^2 + L_h \|h\|^2 + L_I \|I\|^2 + L_\rho \|\rho\|^2$$

We introduce the use of a gradient descent (GD) algorithm with block convex criterion (see Figure 1). For each unknown variable ($h$, $\rho$, and $I$), a convex functional is minimized until close estimates for each unknown are found based on a convergence criterion. The convex functional of the object $\rho$ has two fixed parameters, $h$ and $I$. Likewise, the convex functional of the PSF has $\rho$ and $I$ fixed, while the convex functional of the illumination has $\rho$ and $h$ fixed such that:

$$\nabla_\rho J = \nabla_\rho \sum_i \left\| y^i - h^k \otimes (\rho^k . I_i^k) \right\|^2$$
$$\nabla_h J = \nabla_h \sum_i \left\| y^i - h^k \otimes (\rho^k . I_i^k) \right\|^2$$
$$\nabla_I J = \nabla_I \sum_i \left\| y^i - h^k \otimes (\rho^k . I_i^k) \right\|^2$$

In practice, we record a stack of images shifting the object with respect to a speckle pattern illumination. From the set of images, we first estimate the shift positions for each image based on the position of the cross-correlation maxima. The distance of the maxima to the center gives the distance between two images.

For fluorescence imaging, we use the constraint that $h$, $I$, and $\rho$ are semidefinite positive and have domain limited in the Fourier space with respective domains $D_h$, $D_I$, and $D_\rho$. We use a standard diffraction-limited PSF (for a circular aperture) as initial guess for $h$ and we adjust the frequency content, based on the system NA, to improve the performance of the algorithm. For the object $\rho$, we compute the mean image over all the acquisitions, and for the illumination $I$, we generate the initial guess using the average of the raw images. Since the acquired images are fluorescence intensity images, they are proportional to the speckle intensities.

These initial guesses and the shift positions estimates are input to start the GD algorithm. At each iteration, the object, PSF and illumination are updated, and we empirically update the shift positions every 10 iterations. When each convex function converges, a best estimate for the unknowns is reached. In practice, the algorithm does not always converge to a satisfactory solution (local minimum) and can be refined by adjusting the initial parameters.

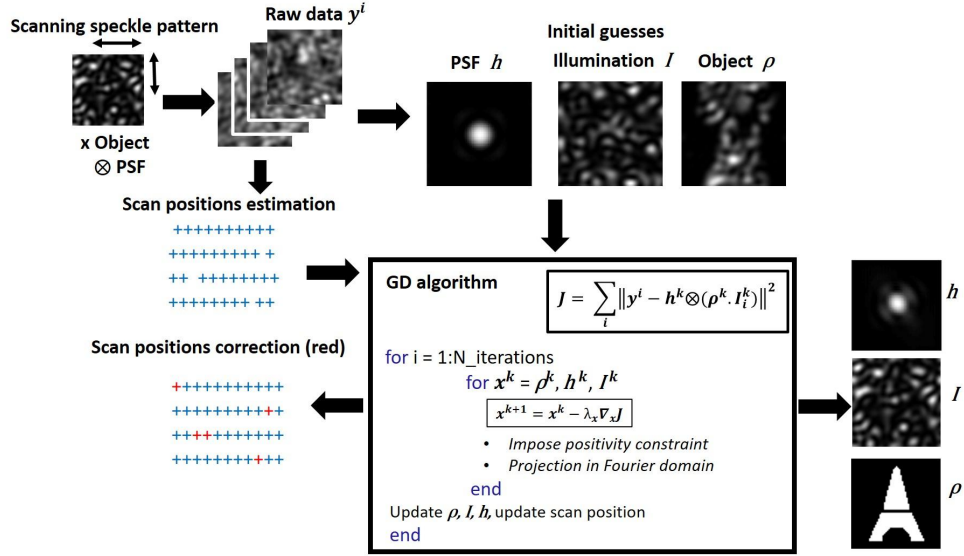

**Fig. 1. Principle of the B-SIFI optimization using a Gradient Descent (GD) algorithm.** According to our model, an object is scanned relative to the illumination pattern and convolved with the PSF of the system. The raw data collected consists of shifted images of the object. From the stack of images, the scan positions are estimated. Initial guesses for the PSF, illumination, and object are created and used as inputs along with the scan position estimations in a GD algorithm. At each iteration the gradient of the convex functions of each unknown (PSF, object, illumination) is calculated with a positivity constraint and low-pass filtered, before the initial guessed values are updated together with the scan positions.

Higher resolution and aberration correction are possible with B-SIFI depending on the system optical parameters. B-SIFI is a SI technique so it can potentially resolve features beyond the diffraction limit from the high frequency speckle illumination. The maximum resolution obtained with SI depends on the different NAs of the imaging system [3]. The imaging system collection diffraction limit is given by $D_{emission} = \frac{0.61 \lambda_{em}}{NA_{sys}}$, where $\lambda_{em}$ is the emission wavelength and $NA_{sys}$ is the imaging system NA. Using an illumination pattern with higher frequency content than the cutoff frequency of the imaging system, translates into a potential increase in resolution of $(NA_{illumin} + NA_{sys})/NA_{sys}$. Hence, in epi-fluorescence microscopy, using a high NA objective maximizes the resolution with linear SI [3], [19]. The resolution will ultimately depend on the illumination NA ($NA_{illumin}$), as well as on the achievable signal to noise ratio (SNR). Practically, if the speckle illumination NA is greater than the system NA, a resolution gain above 2 is expected with respect to the collection diffraction limit. However, this is not always possible in epi-fluorescence microscopy. Here, the focus is on recovering an image limited by diffraction of the collection optics in the presence of aberrations.

*2.1 Simulations*

To test the performance of the B-SIFI algorithm, we run simulations using a Siemens star as the object imaged through an optical system with different aberrations and noise levels (see supplementary information S1). The reconstructed image using the algorithm described in the previous section is compared with the wide-field image that would be obtained using the aberrated system. We use the Structural Similarity Index (SSIM) (ssimval built-in function in Matlab) to have a quantitative value of the object reconstruction. Figure 2 shows an example of the results using a system with a strong coma aberration (Zernike coefficient $Z_3^{-1} = 0.1$) where the algorithm recovers the PSF, the illumination, and the object with good fidelity (SSIM=0.64).

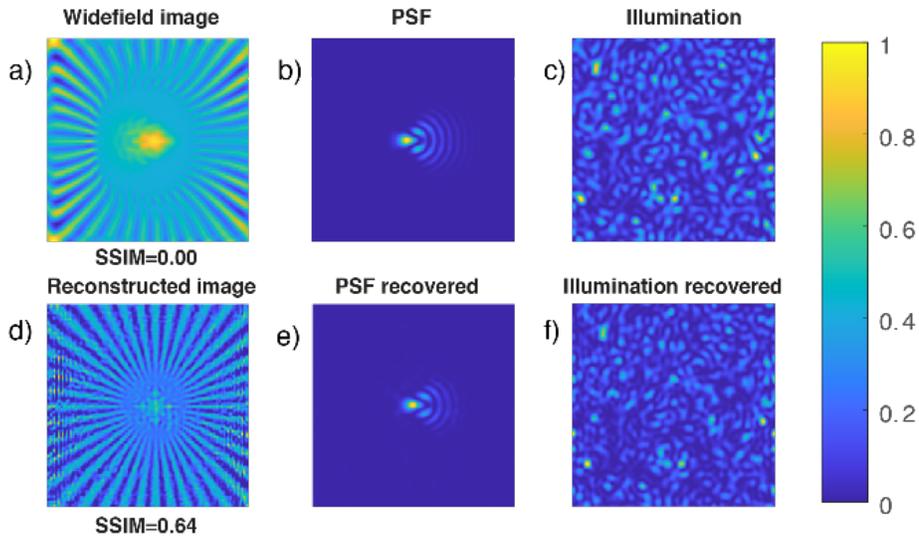

**Fig. 2. Simulation reconstruction using the B-SIFI algorithm without noise**. First row: simulated wide-field image with an aberrated system a), the corresponding aberrated PSF b), and the speckle illumination c) used for the simulation. Second row: Reconstructed object d), recovered PSF (e), and recovered illumination f), using the B-SIFI algorithm.

## 3. Experimental results

We demonstrate the performance of B-SIFI in a custom-built microscope with simple plano-convex lenses. Such a system has inherent aberrations and the images obtained are not diffraction limited. The set-up is represented in Figure 3. A He-Ne laser is spatially filtered and then expanded using a 4f system (f1=300mm, f2=75.6mm) illuminating an engineered diffuser (Newport, angle=10°, 10DKIT-CI) used to generate a controlled static speckle. The beam then goes through a dichroic beam splitter (Semrock 635nm, laser BrightLine single-edge laser Di02-R635) which separates the illumination from the emitted fluorescence. The objective lens L3 is a plano-convex lens of focal length f3=18mm, with a diameter of 6mm and nominal NA=0.17. The measured speckle grain size in the object plane is 2.6μm that corresponds to an effective NA of ~0.12 (see supplementary information S3 for speckle size details using different diffusers). The emitted fluorescence is collected by the same L3 lens, passing through a band-pass filter (Semrock BLP01-635R-25 Longpass filter 633nm EdgeBasic) and is imaged on an EMCCD (Andor iXon 3) using lens L4 (f4=500mm). To simplify the experimental setup, we translate the object while the illumination is static, instead of scanning the illumination while the object does not move. We use two compact motorized translation stages (Thorlabs MTS25-Z8) to translate the object across the field of view.

Additionally, we place a layer of plastic film behind lens L3 to produce a random external aberration in the system. We acquired data both with and without this external aberration, to compare the performance of the algorithm in both cases (see supplementary information S2).

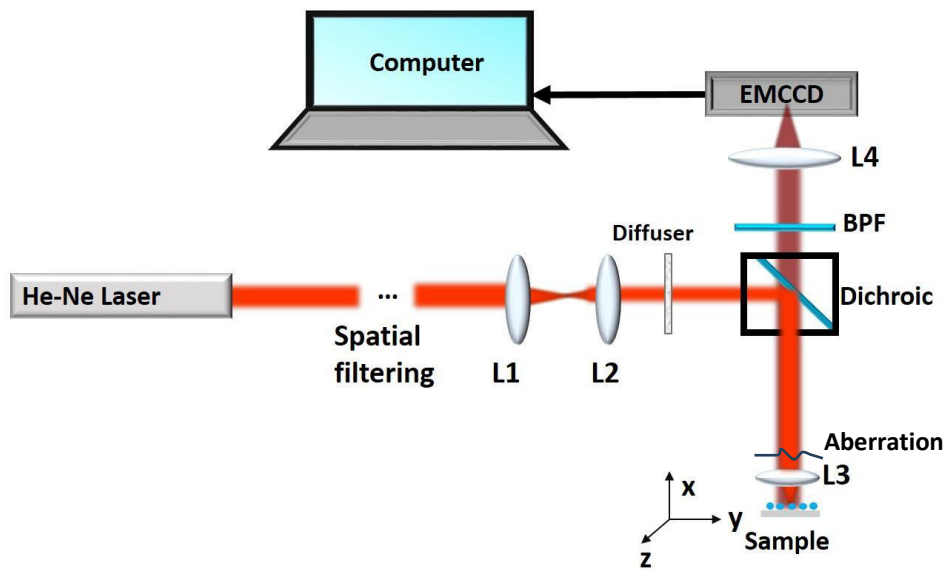

**Fig. 3. Experimental set-up used to demonstrate B-SIFI**. A He-Ne Laser beam is spatially filtered and resized via lenses L1 and L2 (focal lengths 300mm and 75.6mm) before going through an optical diffuser and being focused onto the sample, represented as fluorescent beads here (Sky Blue 2μm, Spherotech). The objective lens L3 has a focal length of 18mm, diameter of 6mm and NA of 0.17. A plastic film is placed adjacent to L3 to induce a random aberration in the system. The emitted fluorescence passes through a band-pass filter (Semrock BLP01-635R-25 Longpass filter 633nm EdgeBasic) and is imaged on an EMCCD (Andor iXon3) with L4 (f4=500mm). The sample is translated in the x-y plane using compact motorized translation stages.

Figure 4 a) shows the fluorescent (wide-field) image of a set of beads (fluorescent Sky Blue beads, Spherotech 2μm) imaged using the setup of Figure 3 under uniform illumination (where the diffuser is replaced by a lens of 100mm focal length). Because of the low NA of the imaging lens and the inherent aberrations, the beads are blurry and not resolved.

To obtain an image of beads using the B-SIFI technique, we scan the object in a 30x30 square grid at 1μm steps recording a total of 900 images. Figure 4 b) shows the resulting image of the beads after 1100 iterations of the reconstruction algorithm and the recovered speckle illumination is shown in Figure 4 c). We plot the 2D Fourier transform of each image in a logarithmic scale to estimate the frequency content of each image and the corresponding NA (Fig. 4 d-f). The frequency content validates the expected improvement in overall system bandwidth based on the use of structured illumination. The PSF given by the algorithm shows the aberration of the optical system, as shown in Fig. 4 g) with a cross section superimposed in magenta. The actual shift positions of the images are also recovered in good agreement with the ground truth. Figure 4 h) shows the initial shift position estimate in blue and the correction made by the algorithm in red, close to a perfect 30x30 linearly spaced grid. The improvement over the wide-field image confirms the potential of this method to correct for aberrations in fluorescence microscopy.

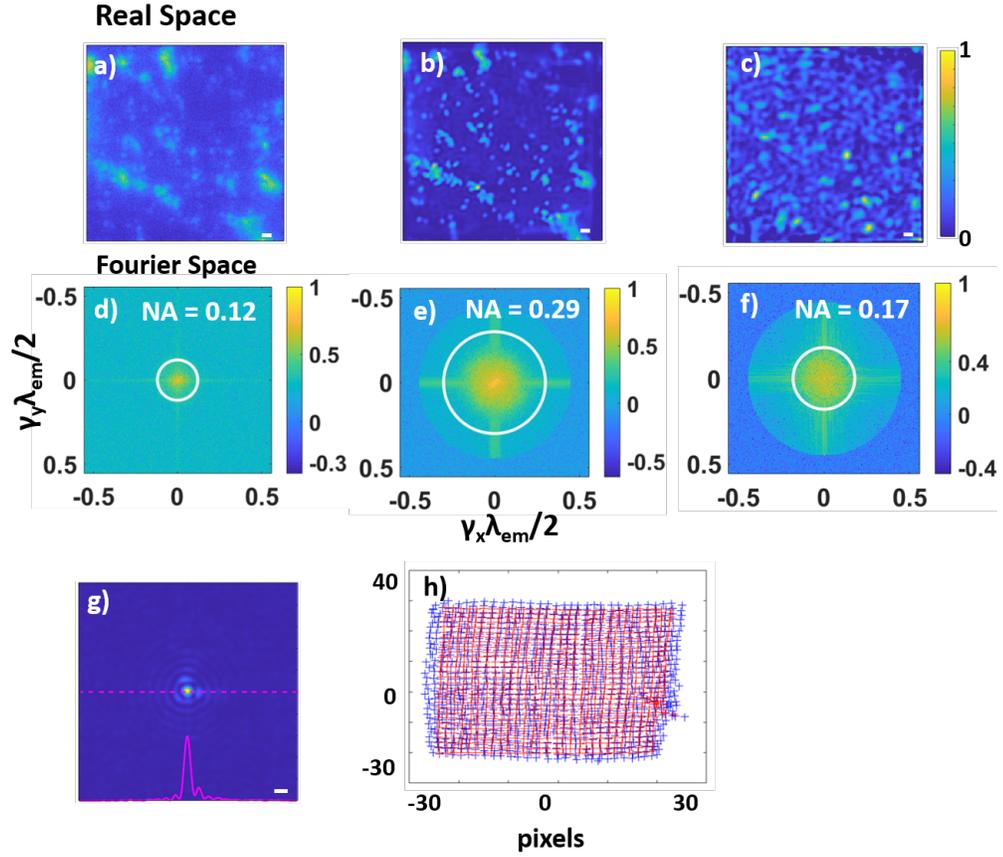

**Fig. 4. Experimental results of B-SIFI in a sample with 2μm fluorescent beads.** a) Wide field image of the object with external aberration. b) Reconstructed image of the object after 1100 iterations of the GD algorithm. c) Reconstructed illumination pattern. Scale bar is 2μm. d) Fourier space representation of the wide field image of the object, corresponding to an NA of ~0.12. e) Fourier space representation of the reconstructed image of the object. The NA circle is the theoretical estimate corresponding to the sum of the system NA and illumination NA. f) Fourier space representation of the reconstructed illumination pattern, corresponding to an NA of ~0.17. g) Reconstructed PSF of the aberrated system. h) Estimate and recovery of the scan positions in blue and red respectively. The axes are in pixels.

To enable quantification of these results in more detail, Figure 5 displays a zoomed area with a few beads. The image of the beads under uniform illumination (Fig. 5 b, f) is compared to the shift corrected average intensity image (Fig. 5 c, g) with the values used as initial guess in the GD algorithm, and to the reconstructed images using B-SIFI (Fig. 5 d, h). Figure 5 e) and i) are plots of the normalized intensity cross sections. The achieved resolution is ~ 2μm. The expected diffraction-limited resolution with uniform illumination is $D_{emission} = \frac{0.61\lambda_{em}}{NA_{sys}} = \frac{0.61*700nm}{0.17} = 2.5\mu m$. Hence the experiments, in addition to correcting for the aberrations, show a resolution improvement as a result of the SI. We note that SI techniques, in the absence of aberrations, can potentially allow for a higher resolution improvement of a factor 2.

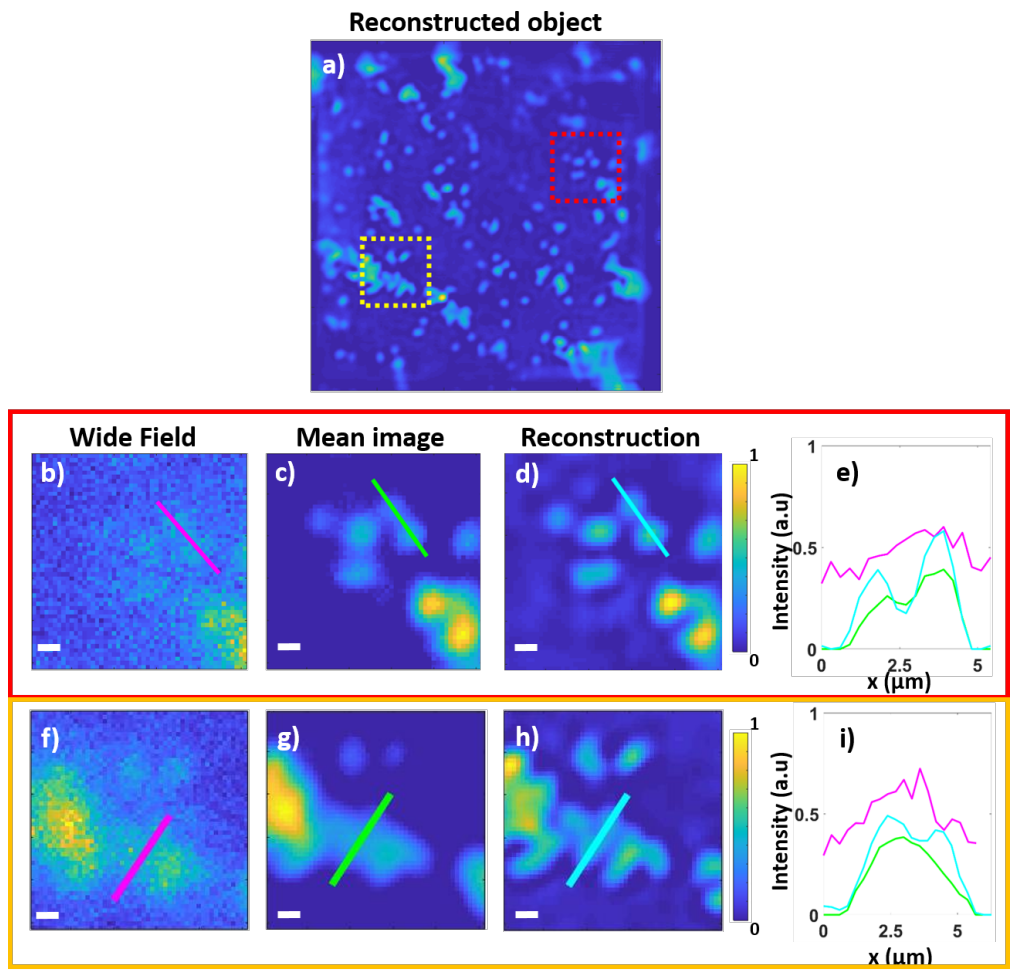

**Fig. 5. Detail of experimental results of B-SIFI** a) Reconstructed object after 1100 iterations of the GD algorithm, the red and orange areas are shown below: b) and f) wide field image of the object, c) and g) mean image of the object illuminated with a scanning speckle pattern, d) and h) reconstructed object. Scale bar is 2 $\mu$m. The cross sections represented in pink, green and blue are plotted in e) and i) in the same colors.

To validate the potential use of the B-SIFI approach with a biological sample, we imaged a thick mouse brain slice dyed with Cy5 using the same system shown in Figure 3. No external aberrations were introduced for these experiments, with the only aberrations being the ones inherent to the system and sample. As illustrated in Figure 6, the object was successfully reconstructed with B-SIFI (Fig. 6 e), showing a significant gain in resolution over the wide field image (Fig. 6 c). Figure 6 e) shows the B-SIFI image after 1040 iterations of the reconstruction algorithm. To verify that the actual object was reconstructed and not an artifact, we also imaged the sample with a high resolution commercial wide-field microscope (Nikon, 40X 0.95NA objective). Figure 6 a) shows the sum of ten images taken with the 40X objective through different z planes, 1μm apart. Note that the objective lens used in our custom set-up has a longer depth of focus than the high-end microscope objective, which provides information over a deeper volume. Thanks to B-SIFI, it is possible to extract high resolution images from an objective lens with such a long depth of focus, which is an advantage in many experimental situations.

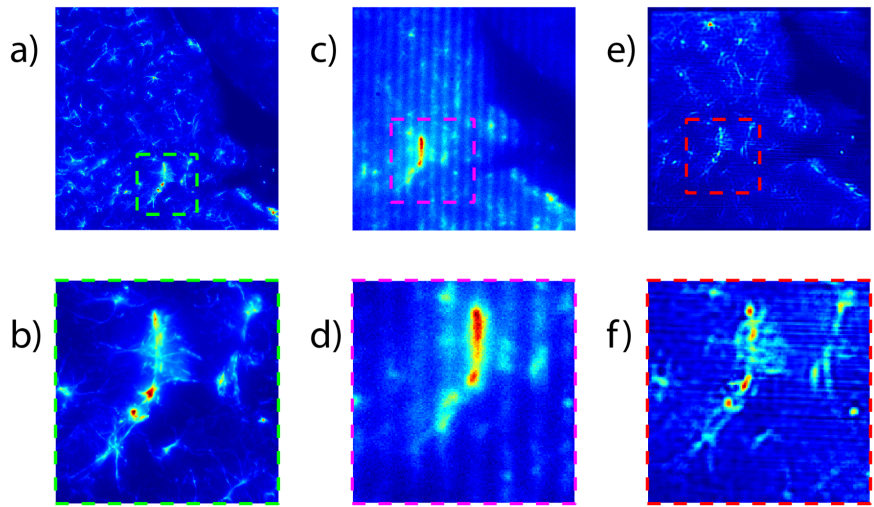

**Fig. 6. B-SIFI imaging of a biological sample** (mouse brain slice dyed with Cy5). a) Sum of 10 wide-field images taken with a 40X 0.95NA microscope objective through different z planes 1μm apart and b) corresponding zoomed area. c) Wide-field image taken with the system EMCCD camera without SI, and d) corresponding zoomed area. e) Reconstructed image with B-SIFI and f) the corresponding zoomed area where the shape of the neuron matches b). Note the gain in resolution from d) to f) due to the use of B-SIFI.

## 4. Discussion and Conclusion

We have demonstrated the B-SIFI imaging approach to improve the resolution of fluorescence microscopy, effectively achieving beyond diffraction-limited resolution while recovering the system's aberrations. There are key advantages to this technique: 1) It does not require any a priori knowledge of the system aberrations in order to reconstruct a high-resolution image; 2) It is easily adaptable to most epi-fluorescence microscopes, making it an affordable solution compared to more sophisticated aberration correction systems [20].

In the experiments above, the resolution could be further improved with smaller speckles in the illumination increasing the effective illumination NA. B-SIFI can be applied to traditional epi-fluorescence microscopy set-ups or transmission geometry set-ups. As pointed above, B-SIFI works whether the object or the speckle illumination is translated, making it adaptable to different experimental conditions.

In conclusion, we presented the B-SIFI imaging technique to reconstruct high resolution, aberration corrected epi-fluorescence images without requiring knowledge of the PSF or speckle illumination. The experimental requirement is simple and the technique can be adapted to various fluorescence microscopy set-ups. The algorithm also reconstructs the speckle illumination and the PSF of the system that encodes the aberrations. The B-SIFI approach is hence particularly attractive in constrained imaging environments, such as the eye [21]. For example, the ocular fundus contains fluorescent structures that can provide useful high-resolution information [22] if the unknown aberrations can be overcome.

**Funding.** National Institutes of Health (NEI/R21 EY029584); NSF award # 1548924.

**Acknowledgments.** We thank Dr. Joseph Dragavon and Dr. Jian Wei Tay for providing the biological sample and for their help with the high resolution images taken at the Advanced Light Microscopy Core at the BioFrontiers Institute, Boulder, Colorado.

**Supplemental document.** See Supplement 1 for supporting content.

**References**

1. M. Laslandes, M. Salas, C. K. Hitzenberger, and M. Pircher, "Increasing the field of view of adaptive optics scanning laser ophthalmoscopy," *Biomed. Opt. Express* **8**(11), 4811–4826 (2017).

2. X. Zhou, P. Bedggood, and A. Metha, "Limitations to adaptive optics image quality in rodent eyes," *Biomed. Opt. Express*. **3**(8), 1811–1824 (2012).

3. M. G. L. Gustafsson, "Surpassing the lateral resolution limit by a factor of two using structured illumination microscopy," *Journal of Microscopy* **198**(2), 82–87 (2000).

4. M. G. L. Gustafsson, "Nonlinear structured-illumination microscopy: Wide-field fluorescence imaging with theoretically unlimited resolution," *PNAS* **102**(37), 13081–13086 (2005).

5. M. G. L. Gustafsson, *et al.*, "Three-dimensional resolution doubling in wide-field fluorescence microscopy by structured illumination," *Biophysical Journal* **94**(12), 4957–4970 (2008).

6. L.-H. Yeh, L. Tian, and L. Waller, "Structured illumination microscopy with unknown patterns and a statistical prior," *Biomed. Opt. Express* **8**(2), 695–711 (2017).

7. J. W. Goodman, "Speckle Phenomena in Optics: Theory and Applications," *J Stat Phys* **130**(2). 413–414 (2008).

8. J. García, Z. Zalevsky, and D. Fixler, "Synthetic aperture superresolution by speckle pattern projection," *Opt. Express* **13**(16), 6073–6078 (2005).

9. J. C. Dainty, Ed., "Laser Speckle and Related Phenomena," vol. 9. in *Topics in Applied Physics* (Springer, 1975), Vol. 9.

10. H. Yilmaz, E. G. van Putten, J. Bertolotti, A. Lagendijk, W. L. Vos, and A. P. Mosk, "Speckle correlation resolution enhancement of wide-field fluorescence imaging," *Optica* **2**(5), 424–429 (2015).

11. E. Mudry, *et al.*, "Structured illumination microscopy using unknown speckle patterns," *Nature Photonics*, vol. **6**(5), 312–315 (2012).

12. S. Labouesse, *et al.*, "Joint reconstruction strategy for structured illumination microscopy with unknown illuminations," *IEEE Trans. on Image Process.*, vol. **26**(5), 2480–2493 (2017).

13. S. Labouesse, *et al.*, "Blind fluorescence structured illumination microscopy: A new reconstruction strategy," in *2016 IEEE International Conference on Image Processing (ICIP)* (2016), pp. 3166–3170.

14. K. Guo, *et al.*, "13-fold resolution gain through turbid layer via translated unknown speckle illumination," *Biomed. Opt. Express* **9**(1), 260–275 (2018).

15. H. Zhang, *et al.*, "Near-field Fourier ptychography: Super-resolution phase retrieval via speckle illumination," *Optics Express* **27**(5), 7498–7512 (2019).

16. L.-H. Yeh, S. Chowdhury, and L. Waller, "Computational structured illumination for high-content fluorescence and phase microscopy," *Biomed. Opt. Express* **10**(4), 1978–1998 (2019).

17. L.-H. Yeh, S. Chowdhury, N. A. Repina, L. Waller, and L. Waller, "Speckle-structured illumination for 3D phase and fluorescence computational microscopy," *Biomed. Opt. Express* **10**(7), 3635–3653 (2019).

18. M. Leonetti, A. Grimaldi, S. Ghirga, G. Ruocco, and G. Antonacci, "Scattering Assisted Imaging," *Scientific Reports* **9**(1), 1–7 (2019).

19. S. Chowdhury, W. J. Eldridge, A. Wax, and J. A. Izatt, "Structured illumination multimodal 3D-resolved quantitative phase and fluorescence sub-diffraction microscopy," *Biomed. Opt. Express* **8**(5), 2496–2518 (2017).

20. J. Chung, G. W. Martinez, K. C. Lencioni, S. R. Sadda, and C. Yang, "Computational aberration compensation by coded-aperture-based correction of aberration obtained from optical Fourier coding and blur estimation," *Optica* **6**(5), 647–661 (2019).

21. Y. Lai-Tim, *et al.*, "Super-resolution *in vivo* retinal imaging using structured illumination ophthalmoscopy," arXiv:2007.16028 (2020).

22. K. Grieve, *et al.*, "*In vivo* near-infrared autofluorescence imaging of retinal pigment epithelial cells with 757 nm excitation," Biomed. Opt. Express **5**(12), 5946–5961 (2018).